\documentclass[aps,prl,superscriptaddress,nofootinbib,twocolumn]{revtex4}

\usepackage{graphicx}
\usepackage{dcolumn}
\usepackage{bm}
\usepackage{wrapfig}

\newcommand{\gese}{Ge$_{x}$Se$_{1-x}$ }

\begin{document}

\title{Electronic Signature of the Intermediate Phase in \gese Glasses:\\
 A Joint Theoretical and Experimental Study}
\author{F. Inam}
\affiliation{Dept. of Physics and Astronomy, Ohio University, Athens OH 45701, USA}
\author{G. Chen}
\affiliation{Dept. of Physics and Astronomy, Ohio University, Athens OH 45701, USA}
\author{D. N. Tafen }
\affiliation{Physics Department, West Virginia University, Morgantown, WV 26506, USA}
\author{D. A. Drabold}
\affiliation{Dept. of Physics and Astronomy, Ohio University, Athens OH 45701, USA}

\date{\today}

\begin{abstract}
Raman and calorimetric studies on  \gese glasses have provided evidence for the existence of the intermediate phase (IP) in chalcogenide and other glasses. Here,  we present X-Ray Absorption Near Edge Structure (XANES) measurements on germanium selenide glasses in the IP composition range, and detect an electronic signature of the IP. \emph{Ab initio} molecular dynamics (MD) based models of these glasses are discussed, and an atomistic picture of the IP, based upon the models and available experiments is presented. We show that these models reproduce detailed composition-dependent structure in the XANES measurements, and otherwise appear to properly represent the structure of the GeSe glasses near the IP.

\end{abstract}

\maketitle

Few binary materials enable scientific inquiry across a continuous range of compositions. This is inconceivable in crystalline systems, which, even in the most complex materials, have a finite number of phases available. An alternative is to study amorphous materials, where glass formation is usually possible for continuous composition ranges. An attractive example of such a system is Ge$_x$Se$_{1-x}$. In sustained study of these and other chalcogenides, Boolchand\cite{feng97}\cite{punit01} discovered that various experimental observables display consistent and anomalous behavior through a {\it finite composition window} near the floppy to rigid transition of Thorpe\cite{thorpe83} (also associated with the work of Phillips\cite{phillips79}). This window has now been observed in many glasses, and has come to be known as the Intermediate Phase (IP). The name originates in the idea that within the window, the system is sandwiched between a ``floppy" and stressed rigid phase. Boolchand first observed the IP in Raman studies of tetrahedral breathing modes in Ge$_x$Se$_{1-x}$ glasses for $x \in (0.20,0.25)$.  $^{129}$I M$\ddot{o}$ssbauer  measurements reveal \cite{bresser86}\cite{punit07} that the variation in the bonding arrangement of Ge and Se sites above $x \approx 0.10$ deviates from a chemically ordered continous random network (CRN). The non-random character of the IP implies that some form of ordering or structural correlation is present. In this Letter, we report a new {\it electronic} measurement and show that our {\it ab initio} models of the GeSe glasses reproduce this and other features of the IP, and we describe the spatial correlations responsible for the electronic signature of the IP. Our work shows that MD simulations can provide novel insight into the IP.

In view of the experimental evidence that special spatial correlations exist in the IP, it is natural that researchers have sought out structural manifestations of the IP. Studies on the first sharp diffraction peak (FSDP) of the Ge-Se glasses indicated that the inverse peak position and peak area of the FSDP exhibit the telltale ``flattening" in the IP composition range\cite{sharma05}\cite{wang04}. However, recent work by Shatnawi and co-workers could not confirm these observations \cite{shatnawi08}. The same authors also conducted high-energy x-ray diffraction and extended x-ray absorption fine structure analyses on \gese, but could not find structural fingerprints of the IP in the first atomic shell \cite{shatnawi08}.

Thorpe et al. \cite{thorpe00} advanced the first theoretical picture of the intermediate phase, in which they invoked the notion of ``self-organization" -- the tendency of the networks to develop the spatial correlations necessary to create the observables.  Building upon earlier work on rigidity transitions in glasses by Phillips \cite{phillips79}\cite{phillips81} and Thorpe \cite{thorpe83}, they showed that the inclusion of extra bonding constraints in a floppy network gives rise to a rigid but stress-free bonding network, before the network transforms into a stressed rigid phase.   Later Micoulaut and Phillips \cite{micoulaut03}, by constructing networks of \gese using the size increasing cluster approximation (SICA), showed that such a stress free network can be thermodynamically stable. Chubynsky and coworkers studied rigidity percolation in model systems and demonstrated that a form of IP was possible, and potentially generic near the rigidity percolation threshold\cite{moose}.  These theories have advanced our understanding of the IP. However, none of them offer a quantitative comparison with experiments, and none of them are based on models formed with realistic interatomic interactions. Recently, we have presented first principles MD models of these glasses over a wide composition range including the IP window \cite{inam07}. In these models, the evolution of structural parameters such as mean coordination and concentrations of corner-sharing (CS) and edge-sharing (ES) tetrahedra show a non-monotonic behavior. These parameters deviate from a chemically ordered CRN, and saturate in the composition range which coincides with the IP window, suggesting a link between experiments and the models.  To a significant degree our work supports the picture of Micoulaut and Phillips \cite{micoulaut03}, inferences of Boolchand\cite{punit01},  and suggests the nature of the self-organization described in Ref. \onlinecite{thorpe00}.

\begin{figure}
\resizebox{90mm}{!}{\includegraphics{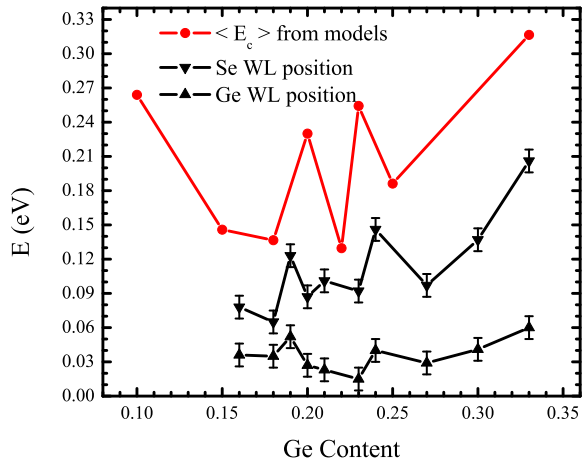}}
  \caption{(Color online) Black points are the K edge WL positions for Se and Ge atoms (error $\pm0.01 eV$) and red points are the average conduction edge energy $\langle E_{c} \rangle$ obtained from the models.}
\label{fig1}
\end{figure}

In this study, we examined several compositions of bulk \gese glasses, with $x=0.16, 0.18, 0.19, 0.20, 0.21, 0.23, 0.24, 0.27, 0.30, 0.33$. The glasses were prepared with a melt-quenching method\cite{wang05}. XANES experiments were performed on beamline 9-BM-B of the Advanced Photon Source at Argonne National Laboratory. Ge K-edge ($E = 11.104 keV$) and Se K-edge ($E= 12.658 keV$) XANES spectra were collected in transmission mode. The bulk glasses were ground into fine powders, which were then spread onto kapton tape. These powders were freshly prepared just before each of the XANES experiment to minimize contact with air. Twelve layers of such tape were stacked to optimize the signal-to-noise ratio of the spectra. A Ge$_{0.33}$Se$_{0.67}$ powder sample was used as a reference to calibrate the absorption edge positions for all the samples.

Strong x-ray absorption peaks are observed at Ge and Se K edges. These two peaks, usually called white lines (WLs), originate from the electronic transition between the core 1s orbital and the lowest unoccupied electronic states (viz., the conduction band). Because of the fully screened core holes, the WL is an indication of the conduction band structure \cite{book}. Any change in the WL peak position implies a relative shift of the electronic states in the conduction band. Fig. \ref{fig1} shows the variation in the relative K-edge WL positions for Se and Ge atoms with Ge content. In case of Se, the WL position decreases to a 'plateau' for $x< 0.33$ with a sudden increase at $x=0.24$ and $x=0.19$ close to the IP window. A similar profile is observed for the WL positions for Ge atoms, though the shift is small compared to the shift in WL position for Se atoms. Note that the Ge K-edge and Se K-edge XANES spectra were obtained independently, so the observed shift in WL positions at $x = 0.19$ and $x=0.24$ is not an artifact of the experiment. Since we have used the same reference for all the samples, the uncertainty in the WL positions was determined by combining the errors from repeated measurements of the same sample as well as from measurements of different samples with the same composition.

\begin{figure*}
\includegraphics[width=6.0 in]{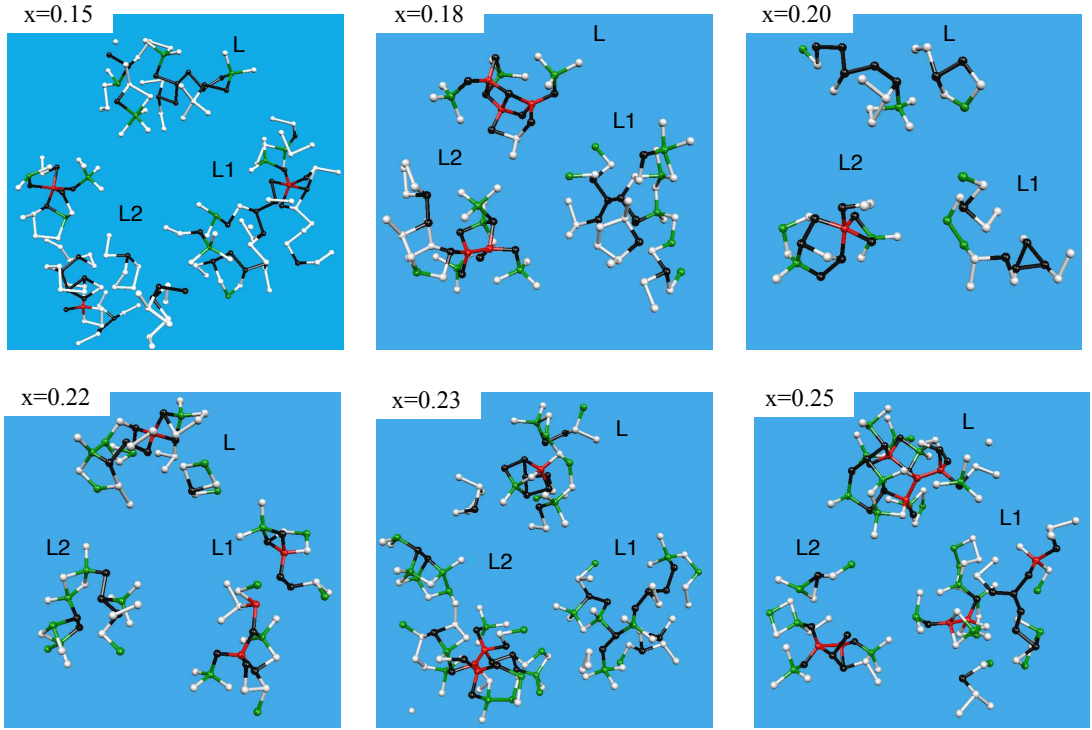}
  \caption{(color online) Conduction edge electron eigenstates in \gese. Black and red sites are Se and Ge atoms respectively which contribute to the lowest three eigenstates (L, L1 and L2, where L stands for ``lowest unoccupied molecular orbital" state) -- states at the conduction tail, and white and green are the neighboring Se and Ge atoms.}
\label{fig2}
\end{figure*}

Atomistic models of the IP were obtained using the approximate {\it ab initio} density functional code FIREBALL \cite{fireball}.  We generated a sequence of 500-atom models with $x =  0.15, 0.18, 0.22, 0.23, 0.25$. Atoms were randomly placed in a cubic cell with fixed volume and experimental density, and equilibrated above the melting point at 1500 K for about 3.5 ps. Then they were cooled to 400K over 4.5 ps, using velocity rescaling. Finally they were steepest-descent quenched to obtain relaxed conformations. The models reproduce the measured pair-correlation functions obtained from X-ray diffraction\cite{inam07}.  Models for $x = 0.10, 0.20, 0.33$ are from Refs. \onlinecite{tafen05,cobb96}.

The shift in the WL positions obtained from XANES can be directly linked to the shift in the conduction band obtained from the models. In Fig. \ref{fig1} red points are obtained by calculating the average conduction edge energy $\langle E_{c} \rangle = \int_{cb} \epsilon \rho(\epsilon)d\epsilon$, where $\rho(\epsilon)$ is the electronic density of states and $cb$ specifies a quadrature range starting at the band edge, extending $1.5$ eV into the conduction states\cite{para}. The absolute values are shifted by a constant for comparison. The similarity between the predicted conduction band energies and the experiments is pleasing: not only the relative shift in $\langle E_{c}\rangle$ is comparable to the experiment but it also shows a sudden increase at $x=0.2$ and $x=0.23$, a feature which is quite clear in the experiment as well.  A small discrepancy in the positions of the two peaks from that of experiment could be due to the small size and rapid quench rates used in the preparation of these models. Apart from that, the overall profile of the shift in the absorption edge with the Ge content is clearly reproduced in the atomistic models.

To elucidate the spatial character of the localized states at the conduction tail, the three lowest eigenstates at the conduction tail for compositions around the IP window are shown in Fig. \ref{fig2}.  It can easily be observed that the reduction of the $\langle E_{c}\rangle$ for $x>0.10$ is correlated with the increase in the appearance of Ge sites (green) around the sites contributing to the conduction tail states. Clustering of Ge creates short strained Se chains which contribute to the localization of the tail states. The effect has increased for $x=0.18$ which corresponds to further lowering of $\langle E_{c}\rangle$. At $x=0.20$ though, we see less clustering of green sites around the eigenvectors which correlates with the sudden increase in $\langle E_{c}\rangle$ at $x=0.20$. Similarly for $x=0.22$ and $0.25$, the eigenvectors are surrounded by more compact Ge clusters as compared to $x=0.23$ where the clusters are relatively open. Visual inspection shows some indication of a correlation between the shift in the conduction edge energy $\langle E_{c}\rangle$ and the clustering of Ge sites. The smaller variation in Ge WL reflects the more uniform Ge bonding environments relative to Se.

\begin{figure}
\includegraphics[width=3.9 in]{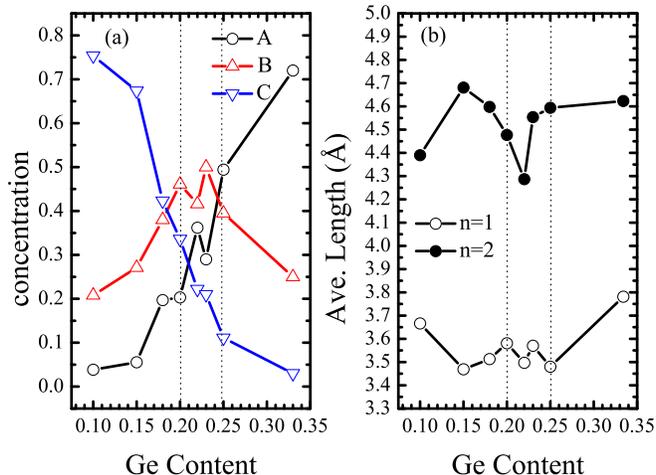}\\
  \caption{a) (Color online) Concentration of sites representing the three regions, which contribute to ten localized states at the valence and conduction tails for different compositions (see text). b) Variation of the average lengths of $n=1$ and $n=2$ chains with the Ge content. Vertical dotted lines indicate the width of the IP.}
\label{fig3}
\end{figure}

To explore these correlations further, we divide the cell in three regions. Region A consists of CS/ES tetrahedra ($n=1$ chains), region B consists of short chains Se$_{n}$ ($n=2,3,4,5$) and region C is composed of longer Se chains representing the amorphous Se background. Fig \ref{fig3}a, shows the concentration of atomic sites belonging to three regions, which contribute to the valence and conduction localized states. In the IP range, tail states are mostly localized in B region, which is understandable as this region appears around the clusters of CS/ES tetrahedra due to cross-linking and thus becomes strained as the size of the clusters and the content of this region increases to a maximum in the IP window. Fig \ref{fig3}b shows the average lengths of $n=1$ and $n=2$ chains, the strain is apparent in terms of the `bending' of these chains. The average length of $n=1$ chain follows a similar profile as the shift in the conduction edge energy (Fig. \ref{fig1}). In order to see the effect of the bending of these chains on the electronic spectrum, we studied a toy model consisting of these two chains attached to each other\cite{chapter}. The decrease in the lengths of these chains shows a clear downward shift in the eigenvalues above the fermi level compared to the relaxed system, which suggests that the antibonding states of this system are more sensitive to the bond angles. The two peaks in the conduction edge energy at $x=0.20$ and $x=0.23$ thus can be understood in terms of the variation in the lengths of these units. The role of filaments in tail states has recently been discussed\cite{urbach}.

A comprehensive analysis of the topological evolution with the Ge content will be presented elsewhere \cite{chapter}. In compact form, addition of Ge in a-Se network nucleates clusters of CS/ES tetrahedra (region A), which can be considered as harbingers of g-GeSe$_{2}$ since CS/ES tetrahedra are the building blocks of this network. Due to the cross-linking at Ge sites, these clusters are surrounded by short Se$_{n}$ chains (region B). Increase in cluster size  also increases the concentration of Se$_{n}$ chains at the cost of the background a-Se (region C). Region B acts as a `barrier' against the formation of GeSe$_2$ (volumes of type A), thus resisting the percolation of region A. The IP appears when most of a-Se background ``C" is transformed into short chains ``B" and the system has many subcritical fragments of g-GeSe$_{2}$ phases and the remainder is short Se$_{n}$ chains. Appearance of such a mixed phase is the cause of the apparent delay in the evolution of the network \cite{inam07}\cite{chapter}, characterized in various experiments as a flattening of observables for $x$ in the IP window. Further increase in the Ge content causes cross-linking in region B, and thus region A percolates all over the system. It seems that the evolution of the region B defines the width and the position of the IP on the composition scale. As Ge atoms tend to form clusters rather then getting dispersed, it ensures the persistence of region B in the IP window. Clustering of Ge sites can be inferred from the increase in the size of FSDP with Ge content \cite{shatnawi08} which is mainly contributed by Ge-Ge correlations \cite{salmon07}. The appearance of short Se chains is apparent from the study of the evolution of the Raman Se chain mode which splits into two modes above $x=0.19$ \cite{salvanathan00}. Higher frequency modes are attributed to short Se chains. In this picture the existence of the IP is due to the appearance of short Se chains. Severing these chains would result in the collapse of the IP. This may explain the reduction in the width of the IP due to the addition of I for the $x=0.25$ composition \cite{wang07}. Also since in the IP range the band tail states reside mainly on short Se chains (fig. \ref{fig3}a), the IP would also be sensitive to the light illuminations as micro-Raman studies suggests \cite{punit2001}.

In summary, we have presented an experimental electronic signature of the IP in \gese glasses  from XANES measurements. First principles models reproduce the composition-dependent shift in the conduction edge. The shift is explained in terms of the change in the lengths of short Se$_{n}$ chain units. An atomic picture of the IP is presented with clear links to the experiments.

We wish to strongly acknowledge Prof. P. Boolchand for providing the samples we employed in the XANES work. GC thanks Ohio University Start-up Fund for supporting this work. DAD thanks the NSF for support under Grants DMR 0605890 and DMR-0600073. FI thanks the NSF for a travel grant under DMR-0409588. Use of the APS was supported by the U. S. Department of Energy under Contract No. DE-AC02-06CH11357.


\begin{references}
\bibitem{feng97} Xingwei Feng, W. J. Bresser and P. Boolchand, Phys. Rev. Lett. \textbf{78}, 4422 (1997).
\bibitem{punit01} P. Boolchand, D.G. Georgiev and B. Goodman, J. Optoelectron. Adv. Mater. \textbf{3}, 703 (2001).
\bibitem{thorpe83} M.F. Thorpe, J. Non-Cryst. Solids \textbf{57} 355 (1983).
\bibitem{phillips79} J.C. Phillips, J. Non-Cryst. Solids \textbf{34} 153 (1979).
\bibitem{bresser86} W. Bresser, P. Boolchand and P. Suranyi, Phys. Rev. Lett. \textbf{56}, 2493 (1986).
\bibitem{punit07} P. Boolchand, P. Chen, M. Jin, B. Goodman and W. J. Bresser, Physica B-Condensed Matter \textbf{389}, 18 (2007).
\bibitem{sharma05} D. Sharma, S. Sampath, N. P. Lalla, and A. M. Awasthi, Physica B \textbf{357}, 290 (2005).
\bibitem{wang04} Y. Wang {\it et. al}, J. Non-Cryst. Solids \textbf{337}, 54 (2004).
\bibitem{shatnawi08} M. T. M. Shatnawi {\it et. al}, Phys. Rev. B \textbf{77}, 094134  (2008).
\bibitem{thorpe00} M.F. Thorpe, D.J. Jacobs, M.V. Chubynsky and J.C. Phillips, J. Non-Cryst. Solids \textbf{266-269} 859 (2000).
\bibitem{phillips81} J.C. Phillips, J. Non-Cryst. Solids \textbf{43} 37 (1981).
\bibitem{micoulaut03} M. Micoulaut and  J. C. Phillips, Phys. Rev. B \textbf{67} 104204 (2003).
\bibitem{moose}M. V. Chubynsky, M.-A. Biere and N. Mousseau, Phys. Rev. E {\bf 74} 016116 (2006).
\bibitem{wang05} Fei Wang, S. Mmedov and P. Boolchand, Phys. Rev. B \textbf{71}, 174201 (2005).
\bibitem{book} D.C. Koningsberger and R. Prins, \textit{X-ray absorption: Principles, Applications, Techniques of EXAFS, SEXAFS and XANES} , Wiley,  New York (1988).
\bibitem{inam07} F. Inam {\it et. al}, J. Phys.: Condens. Matter \textbf{19}, 455206 (2007).
\bibitem{fireball} A. A. Demkov, J. Ortega, O. F. Sankey and M. Grumbach, Phys. Rev. B \textbf{52}, 1618 (1995); O. F. Sankey, D. A. Drabold and A. Gibson, Phys. Rev. B \textbf{50} 1376 (1994).
\bibitem{tafen05} D. Tafen and D. A. Drabold,  Phys. Rev. B \textbf{71} 054206 (2005).
\bibitem{cobb96} M. Cobb, D. A. Drabold and R. L. Cappelletti,  Phys. Rev. B \textbf{54} 12162 (1996); {\it ibid.} Phys. Rev. B {\bf 52} 9133 (1995).
\bibitem{para} Results were insensitive near this range; the band shift was ``rigid" for several eigenvalues (see Ref. \onlinecite{inam07}).
\bibitem{urbach}Y. Pan, F. Inam, M. Zhang and D. A. Drabold, Phys. Rev. Lett. {\bf 100} 206403 (2008).
\bibitem{chapter}F. Inam, G. Chen, D. N. Tafen and D. A. Drabold in {\it Rigidity and Boolchand Intermediate Phases in Nanomaterials} (M. Micoulaut and M. Popescu Eds; INOE, Bucharest) To be published; available as arXiv:0806.0795
\bibitem{salmon07} P. S. Salmon, J. Non-Cryst. Solids \textbf{353} 2959 (2007).
\bibitem{salvanathan00} D. Selvanathan, W. J. Bresser and P. Boolchand, Phys. Rev. B \textbf{61} 15061 (2000).
\bibitem{wang07} F. Wang, P. Boolchand, K. A. Jackson and M. Micoulaut, J. Phys.: Condens. Matter \textbf{19} 226201 (2007).
\bibitem{punit2001} P. Boolchand, X. Feng and W. J. Bresser, J. Non-Cryst. Solids \textbf{293-295} 348 (2001).

\end{references}
\end{document}